\documentclass[11pt]{article}
\usepackage[margin=1in]{geometry}
\usepackage[T1]{fontenc}
\usepackage{lmodern}
\usepackage{microtype}
\usepackage{graphicx}
\usepackage{booktabs}
\usepackage{amsmath}
\usepackage[hidelinks]{hyperref}
\usepackage{xcolor}

\title{Trait, Not State:\\
The Durability of Reading Identity in Social Highlighting}
\author{
  Kazuki Nakayashiki \quad Keisuke Watanabe \\[3pt]
  Glasp Inc. \\
  \texttt{kazuki@glasp.co} \quad \texttt{kei@glasp.co} \\[3pt]
  {\small\itshape Co-first authors (equal contribution).}
}
\date{}

\begin{document}
\maketitle

\begin{abstract}
Prior work on a social web highlighter located individuality in \emph{selection} --- which
documents a person chooses to highlight --- but measured it cross-sectionally. We ask the
temporal question: is a reader's selection signature a \emph{trait} or a \emph{state}? We
freeze each reader's first six months of highlighting as a profile and track its own-vs-other
advantage on their later selections at growing gaps (to $24{+}$ months), with negatives drawn
from the same calendar era --- so supply drift cannot masquerade as personal drift --- at a
coarse global level and at a fine level whose negatives and controls come from the reader's
own interest neighborhood; the anchor cell reproduces the prior cross-sectional level
($+0.188$ vs.\ $+0.169$), validating the harness. Four results. Within the same users, the
fine-layer advantage shows no
statistically detectable paired decline at any reported horizon (6--12 month retention $R=1.00$
$[0.85,1.18]$, $n{=}212$; the farthest bin is compatible with a modest decline; the only
contrast whose interval excludes zero is the coarse layer at 12--24 months, ${\approx}13\%$).
The signal
is not reducible to repeated domains (${\sim}90\%$ survives excluding all profile sources).
Within-person drift is slow (a recent-half profile beats the old half by $+0.042$).
Prospectively, personal profiles --- even one built from a reader's earliest documents,
median 20 months before evaluation --- rank their next reads at roughly $3\times$ the AP of
every simple non-personal prior tested. We use ``trait'' operationally (a stable
signature under continued engagement); the scope is heavy, long-tenured readers of one
platform, and exposure is not separable from choice.
\end{abstract}

\section{Introduction}

A social web highlighter accumulates two kinds of behavioral signal: where people highlight
within a document (salience) and which documents they choose to engage with at all
(selection). In prior work on this platform we showed that the within-document individual
signal is a whisper while the crowd dominates~\cite{nakayashiki2026salience}; that
individuality instead lives at the selection layer, where a reader's history identifies their
documents with a $+0.13$--$0.17$ advantage over matched others, an effect that is
altitude-invariant and mostly thematic~\cite{nakayashiki2026salience,nakayashiki2026selection};
that within-document reader sub-groups are strong but not demonstrably stable across
documents~\cite{nakayashiki2026factions}; and that the aggregate crowd's salience map is
text-predictable on cold documents~\cite{nakayashiki2026coldstart}.

All of these are snapshots. Every measurement held time fixed and asked \emph{where} the
signal lives; none asked how long it lasts. The temporal question decides what the selection
signal \emph{is}: if the advantage a reader's history confers evaporates in months, selection
individuality is a rolling \emph{state} --- a recency effect that any system must chase with
fresh data. If it persists for years, it is a \emph{trait} --- a durable signature, and
long-horizon interest profiles are a legitimate asset.

The two outcomes have opposite practical readings, and the question is not settled by
intuition: production recommenders default to recency weighting and session
models~\cite{koren2009,hidasi2016,kang2018}, implicitly betting on state, while the
psychometric literature finds digital traces predict stable individual
attributes~\cite{kosinski2013,youyou2015}, suggesting trait. We pre-specified both readings
internally (versioned design documents; see Reproducibility) and let the data pick. Throughout, ``trait'' is used \emph{operationally}: a stable,
person-specific selection signature under continued platform engagement --- a property of the
measured behavior (and of the person embedded in their information environment), not a claim
about an immutable psychological attribute.

\textbf{Design in one paragraph.} From a uniformly sampled cohort of long-tenured readers
(${\geq}60$ clean web documents over ${\geq}12$ months), we freeze each reader's
\emph{first six months} of highlighting as a $K$-document profile ($K$ matched between a
reader and their controls), then measure the profile's own-vs-other average-precision
advantage on the reader's \emph{later} selections, binned by the gap since the profile was
frozen ($0$--$1$, $1$--$3$, $3$--$6$, $6$--$12$, $12$--$24$, $24{+}$ months). Negatives are
\emph{time-matched}: drawn from other readers' selections in the same calendar interval as the
positives, so a changing content supply (zeitgeist) cannot masquerade as personal drift.
Difficulty comes in two pre-specified regimes: \textsc{easy} (negatives and controls from the
global cohort --- a coarse topic-identity layer) and \textsc{hard} (negatives and controls from
the reader's top-25 most profile-similar peers --- the fine identity layer, primary for all
gates). The primary retention readout is \emph{paired}: the same user's advantage at a far bin
minus their own 0--1 month value, immune to the composition shift that makes far bins
over-represent long-tenured users.

\textbf{Findings.} (i) The anchor reproduces the prior cross-sectional community-level
estimate: fine-layer advantage $+0.188$ $[+0.160,+0.216]$ vs.\ $+0.169$ before --- an internal
harness validation rather than an independent replication. (ii) \textbf{No detectable decay}:
paired retention at 6--12 months is $R=1.00$ $[0.85,1.18]$; no horizon up to $24{+}$ months
shows a statistically detectable paired decline in the fine layer (the farthest bin,
$n{=}65$, is compatible with a modest late decline); the only paired contrast whose
interval excludes zero is the coarse layer at 12--24 months ($-0.067$, ${\approx}13\%$). No
half-life bin is observed within two years. (iii) The durable signal is not reducible to repeated sources: removing every profile
domain leaves ${\sim}90\%$ of the advantage on matched cells at every horizon. (iv)
Within-person, there \emph{is} slow drift: a volume-matched profile from the recent half of
history beats the old half by $+0.042$ $[+0.020,+0.064]$ --- yet the old half retains
${\sim}91\%$ of the recent half's score. (v) Prospectively, every personal variant --- a
whole-history profile, a recent profile, and even a profile from the reader's earliest
documents --- ranks their actual next documents at ${\sim}3\times$ the AP of every
non-personal prior tested (lifetime popularity, neighborhood co-reading popularity, recency,
random), clearing an internally pre-specified product gate.

\textbf{Contributions.} (1) To our knowledge the first identity-decay curve for naturalistic,
engagement-verified reading selection: a frozen six-month profile keeps its own-vs-other
advantage without statistically detectable loss for at least a year, and through two years
within our resolution. (2) A measurement design that separates personal drift from supply
drift (time-matched negatives), coarse from fine identity (two-regime negatives), and identity
decay from cohort composition (paired retention) --- with the validity failure that motivated
the two-regime design documented transparently. (3) A decomposition showing the durable
identity is not reducible to repeated domains (${\sim}90\%$ survives removing all profile
sources on matched cells) with a small, non-growing habit share. (4) An internally pre-specified
prospective gate connecting the trait result to practice: personal profiles --- even years-old
ones --- are actionable where neither global nor neighborhood popularity priors carry
individual-level information.

\section{Related work}

\textbf{Temporal dynamics in recommendation.} That user preferences drift and that models
should weight recency is a recommender-systems staple: temporal collaborative
filtering~\cite{koren2009}, session-based recurrent models~\cite{hidasi2016}, and
self-attentive sequential recommenders~\cite{kang2018} all encode time. That literature
optimizes next-item accuracy; it rarely measures how long an \emph{identity} signal (one
person's history distinguishing their behavior from a matched peer's) survives, on
engagement-verified traces rather than clicks. We measure exactly that, with the popularity
and sampled-metric cautions of~\cite{ji2020,krichene2020} shaping our baselines and
within-candidate-set evaluation.

\textbf{Digital traces as stable signatures.} Facebook likes predict stable psychological
attributes~\cite{kosinski2013}, and trace-based judgments can rival human
judges~\cite{youyou2015}. Those results are cross-sectional accuracy claims; ours is a
longitudinal durability claim about a behavioral signature, on reading selection.
Highlighting itself predicts comprehension and interest~\cite{winchell2020}.

\textbf{This program.} Paper~I located individuality in selection rather than
salience~\cite{nakayashiki2026salience}; Paper~II showed the selection signal is
altitude-invariant and that personalizing the salience layer fails~\cite{nakayashiki2026selection};
Paper~III found within-document sub-groups strong but not demonstrably
portable~\cite{nakayashiki2026factions}; Paper~IV predicted the aggregate crowd's salience on
cold documents~\cite{nakayashiki2026coldstart}. This paper adds the time axis to the one layer
where individuality robustly lives.

\section{Data and cohort}

\textbf{Platform and sampling.} We study Glasp, a social web highlighter with over a million
registered user records and millions of highlighted URLs. To avoid skew toward super-users
beyond the qualification itself, the cohort was harvested by \emph{uniform} random scanning of
user records: 191{,}223 records were sampled uniformly at random and counted, and every user
with ${\geq}40$ highlighted documents was audited in full. A user qualifies if, after
exclusions, they have ${\geq}60$ clean web documents spanning ${\geq}12$ months with activity
in ${\geq}8$ distinct months ($\geq 2$ documents each) --- 405 qualifying readers (1 in 472
records), median 133 clean documents each (Table~\ref{tab:cohort}).

\textbf{Selection events and exclusions.} A document's \emph{selection time} for a reader is
the minimum valid per-span highlight timestamp (legacy second-resolution values normalized;
coverage was 100\% on every audited user, and the median gap between this and the document's
mutable \texttt{lastUpdated} field is 0 days). Excluded from timelines: imported documents
(import-tagged spans; imports are stamped at import time and would fake selection bursts),
Kindle and social/video domains, and untagged bulk bursts (${\geq}8$ documents sharing one UTC
minute). Exclusions barely bite for qualifying users (median kept fraction $99.5\%$). A
separate audit found note-taking too rare to study (median ${\approx}0\%$ of spans carry a
note).

\begin{table}[t]
\centering
\small
\begin{tabular}{lr}
\toprule
User records scanned (uniform random) & 191{,}223 \\
\quad $\rightarrow$ qualifying readers (strict cohort) & 405 (1 in 472) \\
Median clean web documents / reader & 133 \\
Eligible for the decay analysis (${\geq}12$ profile docs with span text) & 360 \\
Scored (user $\times$ bin $\times$ regime) cells & 2{,}948 \\
Paired users (B0 \& B6, fine layer) & 212 \\
\bottomrule
\end{tabular}
\caption{Cohort funnel. The strict definition (${\geq}60$ clean docs, ${\geq}12$ months,
${\geq}8$ active months) targets readers whose timelines can support multi-year gaps;
platform-wide, ${\sim}4{,}100$ readers qualify (Wilson 95\% lower bound ${\sim}2{,}100$ from
the sampling rate).}
\label{tab:cohort}
\end{table}

\section{Design}

\textbf{Frozen profile.} For each reader: profile window $=$ their first six calendar months
of clean documents; $K=\min(20,\text{window docs})$ profile documents sampled (readers with
$<12$ skipped; 360 remain). The profile representation is the centroid of the reader's own
highlight-span embeddings (\texttt{text-embedding-3-small} at 512 dimensions, ${\leq}4$ spans
per document). The profile is then \emph{frozen}; only the evaluation era moves --- the
cleanest reading of ``how does a fixed profile age?''.

\textbf{Gap bins and positives.} Evaluation documents are the reader's selections after the
profile window, binned by gap since the window's end: $[0,1)$, $[1,3)$, $[3,6)$, $[6,12)$,
$[12,24)$, $[24,\infty)$ months (B0--B24). A (user, bin) cell requires ${\geq}3$ positives
(cap 10).

\textbf{Time-matched negatives.} Each cell draws 40 negatives from \emph{other} cohort
readers' selections in the \emph{same calendar interval} as that user-bin --- so if the
content supply itself drifts, positives and negatives drift together, and the comparison
remains within-era. Negatives exclude any document the evaluated reader ever highlighted and
any document from a control reader's history (the control-in-negatives exclusion
of~\cite{nakayashiki2026selection}). Cells with fewer than 15 available negatives are
skipped; in practice this floor barely binds --- $98.8\%$ (fine) and $99.8\%$ (coarse) of
scored cells received the full 40 negatives (minimum observed 17), and 14 candidate cells
were skipped.

\textbf{Candidate representation.} All candidates (positives and negatives) are represented by
their \emph{title} embedding only --- identical treatment for both classes, and never derived
from the evaluated reader's own spans, which would leak identity.

\textbf{Own vs.\ other.} own $=$ AP of the reader's profile centroid ranking the candidate
set; other $=$ mean AP of three control readers' profiles (built by the same rule, sampled to
the same $K$) on the identical candidates; the \emph{advantage} is own $-$ other. The three
controls are drawn once per reader (seeded) and held fixed across that reader's bins, so
paired contrasts compare the same instrument against the same controls.

\textbf{Two regimes (amended after a smoke test, before the full run).} A 60-user smoke run
produced an implausibly large advantage (${\approx}+0.50$ at every gap, $3$--$4\times$ the
prior cross-sectional level): with globally drawn negatives the task is mostly ``my topics
vs.\ unrelated topics'', a coarse layer that ceilings the score and could mask decay of the
finer signal. We therefore specified, before any full-cohort run, a second regime and made it
primary: \textsc{hard} draws negatives \emph{and} controls from the reader's interest
neighborhood --- their top-25 most profile-similar cohort peers --- translating Paper~II's
community regime to the temporal setting (and neutralizing language clustering by
construction; the smoke cohort was 83\% Latin-script). \textsc{easy} (global) is retained as
the coarse layer. All pre-specified gates apply to \textsc{hard}.

\textbf{Inference and gates.} Per-bin means carry a 3{,}000-iteration cluster bootstrap by
user. The \emph{primary} retention readout is paired: for users contributing both B0 and B6,
$d=\text{adv}(\text{B6})-\text{adv}(\text{B0})$, with retention $R$ the ratio of paired means
--- immune to the composition shift that makes far bins over-represent long-tenured readers.
Pre-specified: the sanity gate $\text{adv}(\text{B0})\geq+0.05$ (the harness must reproduce
Paper~II's cross-sectional signal or stop); the readout $R\geq0.5$ ``trait-leaning'' vs.\
$R<0.5$ ``state-leaning''; and the first bin with advantage below half of B0 as the half-life
bin. Paired contrasts at B12 and B24 were added post hoc as additive robustness (computed the
same way).

\section{Results}

\subsection{The anchor reproduces the prior level, and the curve does not fall}

\begin{figure}[t]
\centering
\includegraphics[width=0.92\textwidth]{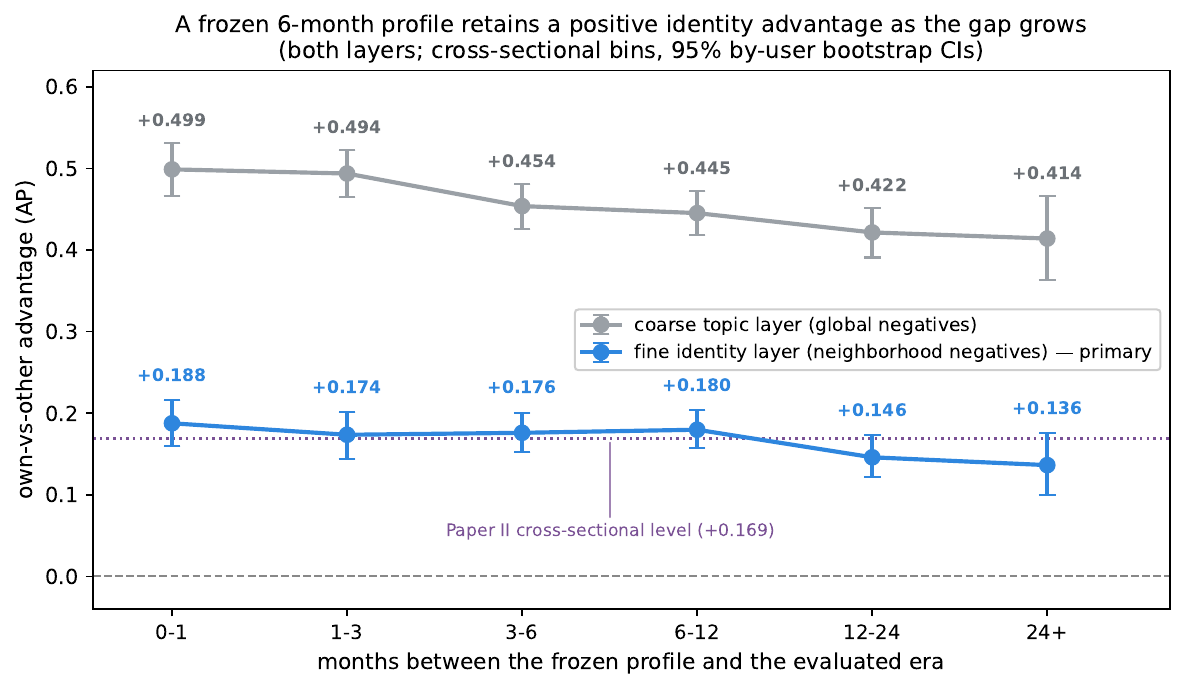}
\caption{The identity decay curve. A profile frozen at month 6 retains a positive
own-vs-other advantage as the gap to the evaluated era grows, at both difficulty layers (cross-sectional
bins; 95\% by-user bootstrap CIs). The fine layer's anchor ($+0.188$) lands on the prior
cross-sectional estimate ($+0.169$, dotted).}
\label{fig:decay}
\end{figure}

\begin{table}[t]
\centering
\small
\begin{tabular}{lcccccc}
\toprule
gap (months) & 0--1 & 1--3 & 3--6 & 6--12 & 12--24 & 24+ \\
\midrule
\textsc{hard} advantage & $+0.188$ & $+0.174$ & $+0.176$ & $+0.180$ & $+0.146$ & $+0.136$ \\
\quad 95\% CI & $[.160,.216]$ & $[.144,.201]$ & $[.152,.201]$ & $[.158,.204]$ & $[.121,.173]$ & $[.100,.176]$ \\
\quad $n$ (users) & 235 & 261 & 306 & 324 & 240 & 102 \\
\quad raw own AP & $.462$ & $.474$ & $.466$ & $.485$ & $.447$ & $.438$ \\
\quad raw control AP & $.274$ & $.301$ & $.290$ & $.305$ & $.301$ & $.301$ \\
\quad held-out advantage & $+0.156$ & $+0.148$ & $+0.149$ & $+0.155$ & $+0.132$ & $+0.126$ \\
\quad\quad 95\% CI & $[.124,.188]$ & $[.117,.179]$ & $[.122,.176]$ & $[.133,.178]$ & $[.106,.159]$ & $[.088,.166]$ \\
\quad held-out/full ratio & $0.93$ & $0.96$ & $0.91$ & $0.91$ & $0.91$ & $0.93$ \\
\midrule
\textsc{easy} advantage & $+0.499$ & $+0.494$ & $+0.454$ & $+0.445$ & $+0.422$ & $+0.414$ \\
\bottomrule
\end{tabular}
\caption{Advantage (own $-$ other AP) by gap bin. Every bin is far above zero
(frac${>}0$, the fraction of by-user bootstrap resamples with positive advantage, is $1.00$
everywhere). Held-out: candidates restricted to domains absent from the
profile; the ratio is computed on the same cells for numerator and denominator (bootstrap CIs
on the ratio span ${\sim}0.84$--$1.02$ across bins). Full per-bin CIs for \textsc{easy} are in
Appendix~A.}
\label{tab:decay}
\end{table}

The fine-layer anchor cell --- evaluation in the month right after the profile freezes ---
gives $+0.188$ $[+0.160,+0.216]$, passing the sanity gate and landing on Paper~II's
cross-sectional community-level estimate ($+0.169$): the harness reproduces the known
cross-sectional result before any temporal claim is made (an \emph{internal} anchor --- same
platform and team, a new harness on different draws --- not an independent replication).
Profile size does not drive the result: fixing $K{=}12$ for \emph{every} reader (rather than
$K=\min(20,\cdot)$) reproduces the picture --- anchor $+0.184$ $[+0.153,+0.214]$, paired
retention $R=0.97$ $[0.82,1.18]$ --- and the advantage is not detectably associated with the
residual variation in $K$ (corr $=0.015$). The result is also not sensitive to the sampled negative draw in this check: all candidate
draws are seeded, and re-running the full pipeline under two alternative sampling seeds
reproduces the picture (anchor $+0.168$ and $+0.191$; paired retention $R=1.08$
$[0.90,1.29]$ and $0.98$ $[0.84,1.15]$). Control calibration behaves:
neighborhood controls score above random on the reader's candidates ($0.27$--$0.30$ vs.\
${\approx}0.20$), as interest-matched controls should, and never below random (the
control-in-negatives pathology of~\cite{nakayashiki2026selection} is absent). Across gap bins
the cross-sectional curve barely moves (Figure~\ref{fig:decay}, Table~\ref{tab:decay}):
$+0.188 \rightarrow +0.180$ at 6--12 months $\rightarrow +0.136$ at $24{+}$ months, every bin
robustly positive. The flatness is not two declines cancelling: within this fixed evaluation harness the
\emph{raw} own AP is itself stable across bins ($0.44$--$0.49$) and so is the control AP
($0.27$--$0.31$; Table~\ref{tab:decay}) --- predictive power is retained in absolute terms
under the same instrument, not merely relative identity. Language subgroups are indistinguishable (all-bins advantage $0.172$
Latin-script vs.\ $0.170$ CJK), confirming the neighborhood regime neutralized that axis.

\subsection{Paired retention: no detectable decay at any reported horizon}

\begin{figure}[t]
\centering
\includegraphics[width=0.8\textwidth]{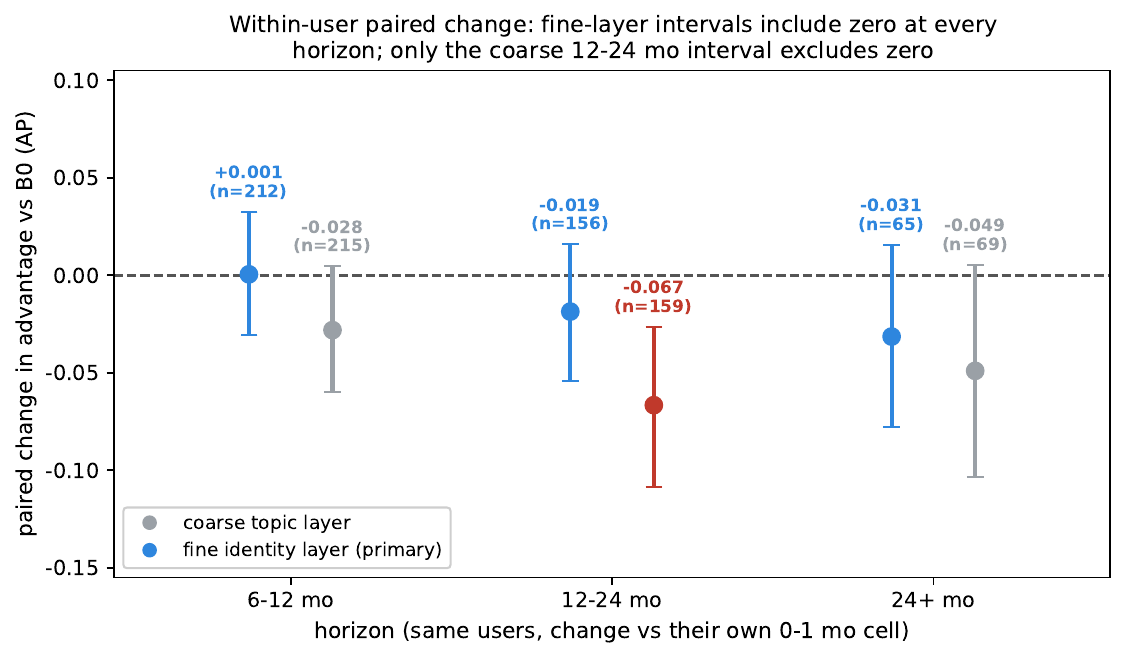}
\caption{Within-user paired change in advantage vs.\ the same user's 0--1 month cell. The
fine identity layer (blue, primary) has an interval including zero at every horizon; the only interval
excluding zero is the coarse layer at 12--24 months (red).}
\label{fig:paired}
\end{figure}

\begin{table}[t]
\centering
\small
\begin{tabular}{llccc}
\toprule
layer & horizon & $n$ & paired $d$ vs.\ B0 & retention $R$ \\
\midrule
\textsc{hard} (fine, primary) & 6--12 mo & 212 & $+0.001$ $[-0.031,+0.032]$ & $1.003$ $[0.854,1.184]$ \\
 & 12--24 mo & 156 & $-0.019$ $[-0.054,+0.016]$ & $0.893$ $[0.705,1.101]$ \\
 & 24+ mo & 65 & $-0.031$ $[-0.078,+0.016]$ & $0.826$ $[0.602,1.093]$ \\
\midrule
\textsc{easy} (coarse) & 6--12 mo & 215 & $-0.028$ $[-0.060,+0.005]$ & $0.944$ $[0.884,1.009]$ \\
 & 12--24 mo & 159 & $\mathbf{-0.067}$ $[-0.108,-0.026]$ & $0.868$ $[0.795,0.942]$ \\
 & 24+ mo & 69 & $-0.049$ $[-0.103,+0.005]$ & $0.897$ $[0.789,1.016]$ \\
\bottomrule
\end{tabular}
\caption{Paired retention: the same users' advantage at a far bin minus their own B0 value
(by-user bootstrap CIs). B6 is the pre-specified primary; B12/B24 are post-hoc additive
contrasts. Bold: the only interval excluding zero.}
\label{tab:paired}
\end{table}

Cross-sectional bins can lie: readers contributing to far bins are the longest-tenured, so a
flat curve could hide real decay behind survivor composition. The paired readout removes this
by comparing each user to themselves (Table~\ref{tab:paired}, Figure~\ref{fig:paired}). The
pre-specified primary --- fine-layer retention at 6--12 months --- is $R=1.003$
$[0.854,1.184]$ ($d=+0.001$ $[-0.031,+0.032]$, $n=212$): no detectable decay, with the
interval excluding losses larger than ${\sim}15\%$. Extending the same contrast post hoc,
the interval includes zero at both 12--24 months ($R=0.893$, $d$ CI $[-0.054,+0.016]$) and
$24{+}$ months ($R=0.826$, $d$ CI $[-0.078,+0.016]$, $n=65$) --- though the $24{+}$ point
estimate is \emph{compatible} with a decline of roughly a sixth that $n{=}65$ cannot
resolve; ``no half-life bin within two years'' is a statement about
what the data show, not proof of indefinite persistence. The per-user paired distribution is
centered on zero and symmetric ($48\%$ of the 212 paired users have $d<0$; median $+0.013$),
so the null average does not mask a sub-population of collapsing readers --- though individual
$d$ estimates are noisy at $\leq$10 positives per cell. The verdict on the pre-specified
readout is \textbf{trait} in the operational sense of Section~1 ($R\geq0.5$ by a wide
margin) --- and it is not specific to the fine regime: in both regimes, the lower bound of
the retention interval remains above $0.5$ at every reported horizon (fine
$0.854/0.705/0.602$, coarse $0.884/0.795/0.789$; Table~\ref{tab:paired}). The only paired
contrast whose bootstrap interval excludes zero (uncorrected, across the six contrasts
examined) is the \emph{coarse} layer at 12--24 months ($-0.067$ $[-0.108,-0.026]$,
${\approx}13\%$ of its level) --- if anything, broad topic mix drifts slightly while the fine
within-neighborhood identity holds.

\subsection{Not reducible to repeated domains}

A reader's durable advantage could simply be ``keeps reading the same sites''. The held-out
arm re-scores every cell with all profile domains excluded from the candidates: on matched
cells (computing full and held-out advantage on the same user-bin cells), the held-out arm
retains $0.91$--$0.96$ of the full advantage at every horizon (bootstrap CIs on the ratio
span ${\sim}0.84$--$1.02$; Table~\ref{tab:decay}). Because held-out scoring drops cells
without enough out-of-profile-domain candidates, the ratio is computed on matched cells and
is \emph{not} the quotient of the two marginal rows in the table. Domain repetition therefore explains only a small
minority of the durable signal, and that share does not grow with the gap. This rules out the
\emph{domain}-level version of source habit; finer source-habit channels (same authors,
newsletters, publication networks, series) are not individually excluded, so the precise
claim is that the durable identity is not reducible to repeated domains and transfers largely
intact to never-before-seen sources.

\subsection{Slow drift within a durable identity}

\begin{figure}[t]
\centering
\includegraphics[width=0.95\textwidth]{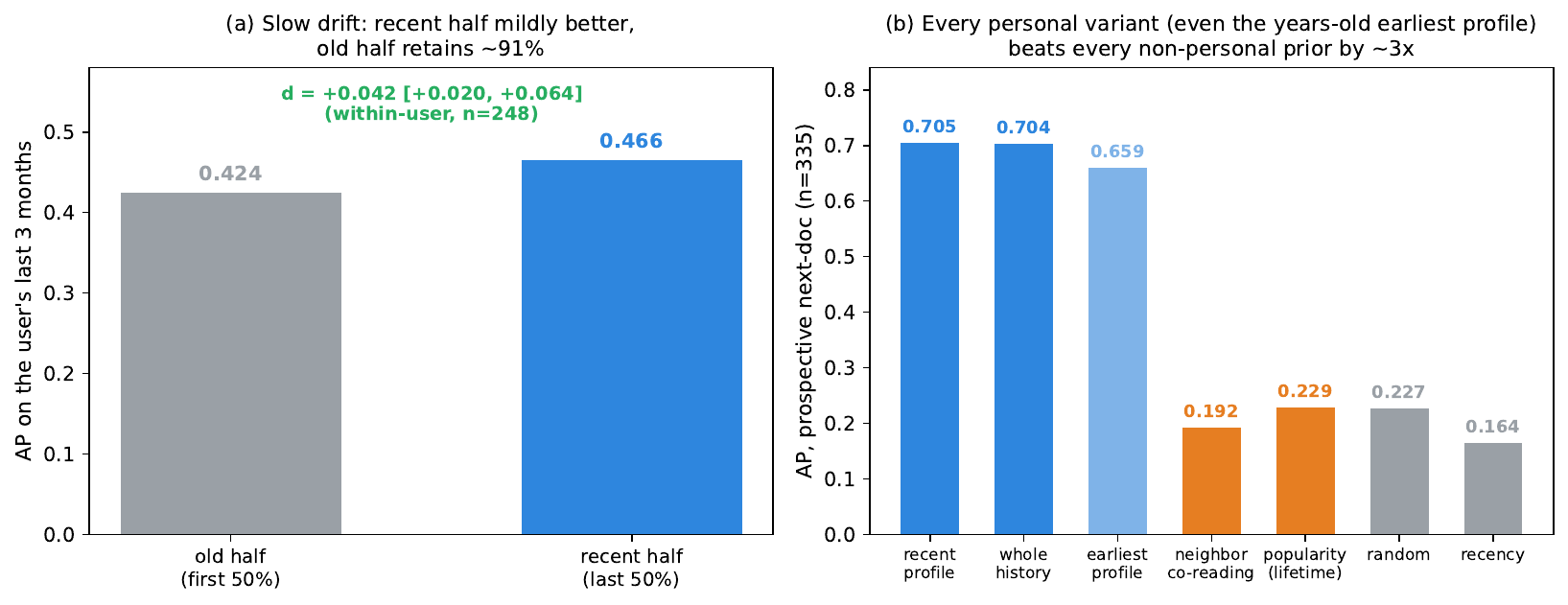}
\caption{(a) Volume-matched halves of the same history: the recent half predicts the
reader's current selections slightly better than the old half ($+0.042$ within-user), but the
old half retains ${\sim}91\%$. (b) Prospective next-doc gate on identical candidates.
Personal variants (recent, whole-history, earliest; blue) vs.\ non-personal priors
(neighborhood co-reading, lifetime popularity, random, recency; orange/gray): every personal
variant far outranks every prior; lifetime popularity ties with random and neighborhood
co-reading falls below it.}
\label{fig:layers}
\end{figure}

Durability against \emph{others} does not mean the person is frozen. Holding volume fixed
($K{=}12$ documents per half), a profile built from the recent half of a reader's history
scores $0.466$ on their current selections vs.\ $0.424$ for the old half: a within-user
difference of $+0.042$ $[+0.020,+0.064]$ (frac${>}0=1.00$, $n=248$;
Figure~\ref{fig:layers}a). Both halves are constructed strictly from documents \emph{before}
the held-out evaluation window (the reader's last three months), so neither profile sees any
evaluated document. There \emph{is} real drift --- but it is slow: the old half alone retains
${\sim}91\%$ of the recent half's score, consistent with the paired decay bounds. The two
results compose into one picture: a person moves a little, but stays far closer to their own
past than to anyone else.

\subsection{Personal profiles are prospectively actionable; simple non-personal priors are not}

Finally, the internally pre-specified product gate: rank each reader's \emph{actual next documents}
(their last three months) inside a time-matched pool of other readers' same-era selections,
comparing rankers on identical candidates (Figure~\ref{fig:layers}b). All personal variants
are built strictly from documents \emph{before} the held-out window. The whole-history
personal profile (20 documents sampled uniformly from the reader's pre-evaluation history,
volume-matched to the earliest/recent variants) reaches $0.704$ AP; the platform's lifetime-popularity prior
reaches $0.229$ --- statistically a tie with random ($0.227$), echoing the popularity-baseline
caveat literature~\cite{ji2020} in reverse: in an individual-level evaluation, a global prior
carries essentially no signal, because individual reading is long-tail. An attempted
collaborative prior --- co-reading count among the reader's own top-25 interest neighbors,
lifetime, with the reader's own documents excluded and ties broken by seeded randomization ---
does no better ($0.192$, at or below random). The diagnostic explains why: only $0.3\%$ of a
reader's actual next reads had been highlighted by \emph{any} of their 25 neighbors, versus
$7.1\%$ of negatives (which are drawn from other readers' selections), so on a long-tail pool
the prior is slightly anti-predictive rather than merely uninformative. (This is a
diagnostic prior under this sampled long-tail candidate construction, not a general verdict
on collaborative filtering.) The gate (personal
$-$ popularity $\geq +0.03$, every resample positive) passes at $+0.475$ $[+0.446,+0.503]$,
and personal $-$ neighborhood-popularity at $+0.512$ $[+0.485,+0.537]$. Echoing the decay
result prospectively: a profile built from the reader's \emph{earliest} 20 documents ---
median $19.6$ months before the evaluation window (IQR $13.5$--$28.7$; $81\%$ at least a year
old, $37\%$ at least two years old) --- still reaches $0.659$ AP, ${\sim}94\%$ of the
recent-20 profile's $0.705$ (recent $-$ earliest $= +0.045$ $[+0.025,+0.065]$, the same
slow-drift magnitude as Section~5.4). The honest scope: the pool is other highlighters'
same-era selections (plausible candidates, not the open web); learned sequential or session
models \cite{hidasi2016,kang2018} are untested here; the gate certifies that the signal is
individually actionable, not that any production system is optimal.

\section{Discussion}

\textbf{What the trait result means in the program.} The program's map of where individuality
lives now has a time axis. Within-document salience: the individual signal is a whisper and
group structure is situational~\cite{nakayashiki2026salience,nakayashiki2026factions}.
Document selection: the individual signal is strong~\cite{nakayashiki2026selection} --- and,
this paper adds, \emph{durable}: it survives years, transfers to unseen sources, and is
prospectively actionable. The asymmetry sharpens: salience-layer structure is shared and
situational; selection-layer structure is personal and persistent. For systems built on this
data the implication is direct: \emph{aggressive forgetting is unwarranted on this signal} ---
durable long-term profiles should be retained and combined with modest recency adaptation
(recency weighting in recommenders~\cite{koren2009,hidasi2016,kang2018} earns its keep on
short-term intent, which our $+0.042$ recent-half edge also shows, not because old preference
data loses identity value). We do not detect a meaningful loss at one year under the
pre-specified retention criterion, and most of the profile's value is not reducible to
repeated domains.

\textbf{Why might the signature be this stable?} What we measure is embedding-level semantic
selection --- the granularity at which Paper~II found individuality thematic --- not deeper
constructs such as reading style, stance, depth, or criticality; ``reading identity''
throughout means this measured selection signature. Tastes at this granularity plausibly
change on the scale of careers and life phases, not months; the small recent-half edge
($+0.042$) is consistent with slow accretion of new themes on a stable base rather than
replacement. A signal invisible to title/span embeddings could decay faster. A second,
non-exclusive reading is environmental: a reader's information environment (follows,
newsletters, communities) is itself persistent and co-evolves with them, so the durable
signature is strictly a property of the \emph{person-in-environment} system. The platform
itself is a browser-extension highlighter over the reader's own web browsing rather than a
feed-ranked surface, which weakens (but does not eliminate) the in-platform
algorithmic-feedback-loop explanation; personalized \emph{external} exposure cannot be
separated without impression logs. Our estimate should therefore be read as the durability of
\emph{observed selection behavior}, not as a pure preference parameter.

\section{Limitations}

\textbf{Survivor cohort.} Qualifying readers are heavy, long-tenured users (1 in 472 records);
durability among light or churned users is unmeasured, and a reader must keep highlighting to
be observable at far gaps --- our claim is conditional on continued engagement, not a claim
that engagement continues. One thing this is \emph{not}, however, is tautological:
qualification selects on sustained \emph{volume and spread} of activity, not on content
coherence --- a reader who completely changed topics every year would qualify just as easily
--- so the measured \emph{content} stability is not implied by the selection rule (though
sustained engagement may itself correlate with interest stability, a weaker selection effect
we cannot exclude).
\textbf{Exposure vs.\ choice (the central alternative).} Without impression logs, selection
mixes what readers were shown with what they chose, so the durable signature is strictly a
property of the person-in-environment system (Section~6). In particular, if recommendation
loops continually re-served early-profile-like content, observed stability could partly
reflect the loop rather than the person; the platform's extension-over-own-browsing structure
and the held-out arm's transfer to never-seen domains weaken a domain-level version of this
account, but a topic-level personalized-exposure loop (external feeds, follows, newsletters)
cannot be excluded. \textbf{Instrument sensitivity.} Profiles and candidates are frozen
title/span embedding centroids --- a deliberately simple, leakage-resistant instrument. A
stronger model could reveal drift this one misses; conversely the measured advantage is a
lower bound on the identity signal at each gap, and the paired design compares the same
instrument across gaps. \textbf{Granularity.} Six gap bins and a 6-month profile window are
coarse; the B24 paired contrast has $n=65$ and a wide interval ($R=0.826$; $d$ CI
$[-0.078,+0.016]$), so a modest late decline is compatible with the data.
\textbf{Negatives define the difficulty.} Advantage levels depend on the negative pool
(coarse vs.\ fine differ by ${\sim}2.7\times$); we therefore anchor on reproduction of the
prior cross-sectional level and interpret \emph{shapes}, primarily paired within-instrument
contrasts. \textbf{One platform.} Glasp readers who highlight for years are a particular
population; the trait/state question deserves replication on other longitudinal reading
traces.

\section{Ethics}

This study analyzes highlighting behavior on a consumer platform under its terms of service,
as internal aggregate-level analytics without experimental intervention. Per-user timelines
were processed only to compute per-user scores; no individual is profiled for any product
decision here, no individual-level result is reported, and documents and readers appear in
released artifacts only as aggregate statistics with confidence intervals. The
extraction/scoring pipeline runs on private data and is not released; highlighting histories
can be identifying, so per-user rows stay internal, and upstream deletion and private-content
requests are honored. A durable-identity result cuts both ways ethically: it supports useful
long-horizon personalization, and it means reading histories retain personal signal for
years --- a reason for conservative retention and access policies, which this finding informs.

\section{Conclusion}

Is a reader's selection signature a trait or a state? On a cohort of long-tenured readers of a
social web highlighter, it behaves like a trait in the operational sense: a profile frozen
after six months keeps its own-vs-other advantage without detectable loss at a 6--12 month gap
($R=1.00$ $[0.85,1.18]$), shows no statistically detectable paired decline through two years
(the farthest horizon remains compatible with a modest decline), transfers largely intact to
never-seen sources (${\sim}90\%$ on matched cells), and prospectively ranks the reader's
actual next documents far above simple non-personal priors, including a
popularity prior that ties with random. Within the person there is slow drift ($+0.042$ for
the recent half of history), but it accretes on a stable base rather than replacing it.
Selection individuality --- the one layer of this program where the individual robustly beats
the crowd --- is not just strong; it lasts.

\section*{Reproducibility}

Every effect size carries a 3{,}000-iteration by-user cluster-bootstrap CI. Pre-specification
is \emph{internal} (we avoid the unqualified term ``pre-registered''): the cohort definition, profile construction, gap bins, regimes, margin,
and pass rules were fixed in design documents versioned in the project repository before each
run, not lodged with an external registry --- we state the boundary precisely so readers can
weight it. The design documents are versioned in the
private monorepo, where the project design and gates predate any run and the step-design
predates the measurement runs; they are available to researchers on reasonable request, and
the pre-specified elements are collected, in condensed form, in Appendix~B. The paper and figures are available at
\url{https://github.com/glasp-co/selection-dynamics}. The two-regime amendment was written into the versioned design document
after the 60-user smoke and before the full-cohort run but was committed together with the
run's results, so its before-run status rests on our statement and the smoke record; at amendment time we had seen only the smoke's aggregate per-bin levels (both
regimes), and no full-cohort number existed --- and the verdict is in any case regime-robust
(Section~5.2), so it does not hinge on the amendment's timing. Future papers in this program
will use an external registry. The paired B6
contrast and its $R\geq0.5$ readout were pre-specified; the paired B12/B24 extensions, the
held-out ratio analysis, the heterogeneity summary, and the added prospective baselines
(neighborhood popularity, earliest/recent profiles) are post-hoc robustness analyses and are
labeled as such where reported. The data-extraction and scoring pipeline runs against Glasp's
private user data and is not released; per-user results derive from individual highlighting
behavior and are not published. The cluster-bootstrap estimator and aggregate statistics are
available to researchers on reasonable request.

\section*{Appendix A: coarse-layer per-bin confidence intervals}

\begin{center}
\small
\begin{tabular}{lcccccc}
\toprule
gap (months) & 0--1 & 1--3 & 3--6 & 6--12 & 12--24 & 24+ \\
\midrule
\textsc{easy} advantage & $+0.499$ & $+0.494$ & $+0.454$ & $+0.445$ & $+0.422$ & $+0.414$ \\
\quad 95\% CI & $[.467,.531]$ & $[.465,.522]$ & $[.425,.481]$ & $[.418,.472]$ & $[.391,.452]$ & $[.364,.466]$ \\
\quad $n$ (users) & 239 & 264 & 307 & 326 & 240 & 104 \\
\bottomrule
\end{tabular}

\smallskip
{\small Table A1: Coarse-layer (\textsc{easy}) advantage by gap bin with 95\% by-user
bootstrap CIs, completing Table~2.}
\end{center}

\section*{Appendix B: pre-specified elements}

Condensed from the internal design documents, as fixed before the corresponding runs (see
Sections~3--4 and Reproducibility for context):

\begin{itemize}\setlength{\itemsep}{1pt}
\item \textbf{Question and readings.} Trait vs.\ state, with both outcomes pre-specified as
reportable findings (flat decay $\rightarrow$ durable profiles; steep decay $\rightarrow$
half-life and recency weighting).
\item \textbf{Cohort gates.} G0a: ${\geq}80\%$ valid per-span timestamps among audited
users, else fall back to document-level timestamps or stop. G0b: ${\geq}60$ clean web
documents, ${\geq}12$ months lifespan, ${\geq}8$ distinct months with ${\geq}2$ documents;
${\geq}200$ qualifying readers platform-wide (Wilson 95\% lower bound), with one
pre-specified fallback (${\geq}40$ documents, ${\geq}9$ months, ${\geq}6$ active months). G0c: qualifying readers keep ${\geq}60\%$ of documents
after import/Kindle/burst exclusions.
\item \textbf{Profile and bins.} First six calendar months; $K=\min(20,\text{window})$,
skip if $<12$; span-embedding centroid. Gap bins $[0,1)$, $[1,3)$, $[3,6)$, $[6,12)$,
$[12,24)$, $[24,\infty)$ months; ${\geq}3$ positives per cell, cap 10.
\item \textbf{Candidates.} 40 time-matched negatives from the same calendar interval
(${\geq}15$ required); exclude the reader's own documents and the controls' documents;
title-only representation, never the evaluated reader's spans.
\item \textbf{Inference and gates.} 3{,}000-iteration by-user cluster bootstrap; sanity
$\text{adv}(\text{B0})\geq+0.05$ else stop; primary readout paired $R=\text{adv(B6)}/
\text{adv(B0)}$ with $R\geq0.5$ trait-leaning, $R<0.5$ state-leaning; half-life = first bin
below $0.5\times$B0. Prospective product gate (Section~5.5): personal $-$ popularity
$\geq\delta=+0.03$ with every resample positive.
\item \textbf{Two-regime amendment} (2026-06-10; after the 60-user smoke, before any
full-cohort run): \textsc{easy} retained as the coarse layer; \textsc{hard}
(interest-neighborhood negatives and controls) added and made primary for all gates. At that
time only the smoke's aggregate per-bin levels were known.
\item \textbf{Labeled post-hoc.} Paired B12/B24 contrasts, the held-out ratio analysis, the
heterogeneity summary, the $K{=}12$ and alternative-seed reruns, and the added prospective
baselines (neighborhood popularity, earliest/recent profiles).
\end{itemize}

\end{document}